\documentclass[twocolumn,showpacs,amsmath,amssymb]{revtex4-1}
\usepackage{chemarr}
\usepackage{etex}
\usepackage{graphicx}
\usepackage{subfigure}
\usepackage{amssymb}
\usepackage{float}

\begin{document}

\author{Marta Galanti$^{1,2,3,4}$, Duccio Fanelli$^{1,3}$, Francesco Piazza$^4$}
\affiliation{$^1$Universit\`{a} degli Studi di Firenze, Dipartimento di Fisica e Astronomia and CSDC, via G. Sansone 1, 
IT-50019 Sesto Fiorentino, Firenze, Italia, $^2$Dipartimento di Sistemi e Informatica, Universit\`a di Firenze, Via S. Marta 3, IT-50139 Florence, Italy,$^3$INFN, Sezione di Firenze, Italia,$^4$Universit\'{e} d'Orl\'{e}ans, Centre de Biophysique Mol\'{e}culaire, CNRS-UPR4301, Rue C. Sadron, 45071, Orl\'{e}ans, France.}

%%\title{Theory of diffusion-reaction chemistry in complex geometries}
\title{Theory of diffusion-influenced reactions in complex geometries}

%%
%% ====================================================================================================================
%%   SELECTED PACS
%%
%%   * 00.	GENERAL
%%   --------------
%%          - 02.	Mathematical methods in physics
%%                   - 02.30.Jr : Partial differential equations
%%
%%   * 80.	INTERDISCIPLINARY PHYSICS AND RELATED AREAS OF SCIENCE AND TECHNOLOGY
%%   -----------------------------------------------------------------------------
%%          - 82.	Physical chemistry and chemical physics
%%                   - 82.40.Ck : Diffusion in  chemical reaction kinetics
%%                   - 82.40.Qt : Reactions in complex chemical systems
%%                   - 82.39.Rt : Reactions in complex biological systems (see also 87.18.-h Biological complexity)
%%    
%%          - 87.	Biological and medical physics
%%                  - 87.10.Ed : Partial differential equations, in mathematical aspects of biological physics
%%
%%=====================================================================================================================
%%


\begin{abstract}
\noindent Chemical reactions involving diffusion of reactants and subsequent chemical fixation steps
are generally termed ``diffusion-influenced'' (DI). Virtually all biochemical processes in living
media can be counted among them, together with those occurring in an ever-growing number of emerging
nano-technologies. The role of the environment's geometry (obstacles, compartmentalization) and
distributed reactivity (competitive reactants, traps) is key in modulating the rate constants of
DI reactions, and is therefore a prime design parameter. Yet, it is a formidable
challenge to build a comprehensive theory able to describe the environment's ``reactive geometry''.
Here we show that such a theory can be built by unfolding this many-body problem through addition theorems
for special functions. Our method is powerful and general and allows one to study a given DI 
reaction occurring in arbitrary ``reactive  landscapes'', made of multiple spherical boundaries 
of given size and reactivity. Importantly, ready-to-use analytical formulas can 
be derived easily in most cases.
\end{abstract}

\maketitle

%%%%%%%%%%%%%%%%%%%%%%%%%%%%%%%%%%%%%%%%%%%%%%%%%%%%%%%%%%%%%%%%%%%%%%%%%%%%%%%%%%%%%%%%%%%%%%%%%%%%%%%%%%
% MAIN BODY 

\noindent Diffusion-influenced reactions (DIR) are ubiquitous in many contexts in physics, chemistry and 
biology~\cite{Rice:1985kx,Szabo:1989fk} and keep on sparking intense theoretical 
and computational activity in many fields~\cite{Foffi:2013fk,Schoneberg:2013aa,Seki:2012ux,
Dorsaz:2010vn,Schmit:2009ij,Ridgway:2008ly,ben-Avraham:2000kx,Traytak:1995,Mitra:1992pt}. 
Modern examples of emerging nanotechnologies that rely on controlled 
alterations of diffusion and reaction pathways in DIRs include different sorts of 
chemical and biochemical catalysis
involving complex nano-reactors~\cite{Lu:2011yq,Welsch:2009ys},
nanopore-based sequencing engines~\cite{Brady:2015aa}
and morphology control and surface functionalization of inorganic-based delivery vehicles
for controlled intracellular drug release~\cite{Gao:2011aa,Vivero-Escoto:2010aa}.\\
\indent However, while the mathematical foundations for the description 
of such problems have been laid nearly a century ago~\cite{Smoluchowski:1916fk},
many present-day problems of the utmost importance at both the fundamental and applied level
are still challenging. Notably, arduous difficulties arise in the quantification of the important 
role played by the environment's {\em geometry}  (obstacles, compartmentalization)~\cite{Benichou:2010}
and {\em distributed reactivity} (patterns of competitive reaction targets or traps) 
in coupling transport and reaction pathways in many natural and artificial 
(bio)chemical reactions~\cite{Shlesinger:1993aa,Kopelman:1988aa,Rice:1985kx}.\\
%
%
% [Vivero-Escoto:2010aa]
%
% morphology control and surface functionalization of inorganic-based delivery vehicles, such as 
% mesoporous silica nanoparticles (MSNs), % have brought new possibilities to this burgeoning area of research. 
% The ability to functionalize the surface of mesoporous-silica-based nanocarriers with stimuli-responsive groups
%
% 
\indent A formidable challenge in modeling environment-related  effects
on chemical reactions is represented by the intrinsic many-body 
nature of the problem. This is brought about essentially 
by two basic features, common to virtually all realistic situations, namely 
($i$) finite density of reactants and other inert species (in biology also referred 
to as {\em macromolecular crowding}~\cite{Ellis:2001wv,Zhou:2008vf}) and 
($ii$) confining geometry of natural or artificial reaction domains in 3$D$ space.
In general, the presence of multiple reactive and non-reactive particles/boundaries
cannot be neglected in the study of (bio)chemical reactions occurring in real {\em milieux},
where the geometrical {\em compactness} of the environment may have profound effects, 
such as first-passage times that are non-trivially influenced by the starting point~\cite{Benichou:2010aa}.    
Relevant complex media include the cell 
cytoplasm~\cite{Kim:2009fk,Dix:2008lp,Ridgway:2008ly,Luby-Phelps:2000zr},
porous or other artificial confining media~\cite{Kurzidim:2011ab,Nguyen:2010fv,Benichou:2010aa,
Kurzidim:2009aa,Novak:2009uq,Talkner:2009,Kim:1992vn},
which be considered as offering important tunable features for technological 
applications~\cite{Brady:2015aa,Lu:2011yq,Vivero-Escoto:2010aa}. \\
\indent In this paper, we take a major step forward by solving the general problem of 
computing the steady-state reaction rate constant for a diffusion-influenced chemical reaction 
between a mobile ligand and an {\em explicit} arbitrary, static 3$D$ configuration of
spherical reactive boundaries of arbitrary sizes and intrinsic reactivities. \\
\indent To set the stage for the forthcoming discussion, 
let us first consider the simple problem of two molecules $A$ and $B$ of size $R$ and $a$, respectively, 
diffusing in solution.  Upon encountering, the two species can form a complex, 
which catalyzes the transformation of species $B$ into some product $P$ with 
rate constant $k$,
\begin{equation}
\label{e:reactionred}
A + B \xrightarrow[]{k} A + P 
\end{equation}
Under the hypotheses that ($i$) $A$ molecules diffuse much more slowly than 
$B$ molecules, ($ii$) both species are highly diluted and ($iii$) the bulk concentration 
of $A$ molecules $\rho_A$ is much smaller than the bulk concentration 
of $B$ molecules $\rho_B$~\cite{Szabo:1989fk,Piazza:2013aa}, the  
rate constant $k$ can be computed by solving the following stationary 
two-body boundary problem~\cite{Rice:1985kx}
%
%
%\begin{subequations}
%\begin{align}
%&\bigtriangledown^2 u = 0\\
%&\left. u \right|_{\partial \Omega_0}=0\\
%&\lim_{r\to \infty} u = 1
%\end{align}
%\end{subequations}
%
\begin{equation}
\label{e:BPSmol}
\bigtriangledown^2 u = 0      \quad \mbox{with} \quad 
\left. u \right|_{\partial \Omega_0}=0,  \,
\lim_{r\to \infty} u = 1
\end{equation}
where $\partial \Omega_0$ is a spherical sink of radius $\sigma = R+a$ 
(the encounter distance) and $u(r) = \rho(r)/\rho_B$
is the stationary normalized concentration of $B$ molecules around the sink. 
The rate constant is simply the total flux into the sink, {\em i.e.}
\begin{equation}
k = D \int\limits_{\partial \Omega_0}
                  \left. \frac{\partial u}{\partial r} \right|_{r=\sigma} dS 
\label{e:ratec}
\end{equation}
where $D=D_A+D_B$ is the relative diffusion constant. 
The solution to the boundary problem~\eqref{e:BPSmol} is $u(r)=1-\sigma/r$, which yields the 
so-called Smoluchowski rate constant for an isolated spherical sink, namely $k_S=4\pi D \sigma$.
%
%    aggiunta come da commento di Giuseppe, che sembrava che la storia morisse un po' li
%
These simple ideas, originally developed to describe coagulation in 
colloidal systems~\cite{Smoluchowski:1916fk,Smoluchowski:1917aa}, together with the related
subsequent major advances by Debye~\cite{Debye:1942aa} and Collins \& Kimball~\cite{Collins:1949aa}
represent the basic building block of many modern theoretical approaches in 
chemical%~\cite{Houston:2001aa} 
and soft-matter~\cite{Muller:2014aa,Oshanin:1994aa} kinetics.\\
%
%   muore un po' cosi ... 
%   citare come campi di stra-applicazione COLLOID aggregation + libro di 
%   + libro text-book verde 
%
%   L'IDEA E' che nell'intro si parla solo di diffusion-limited reactions + ostacoli, geometria...
%
%   General references on concentration dependence of reaction rates + crowding etc
%
%   ~\cite{Gopich:2002uq} 
%
\indent In many realistic situations in chemical and biochemical kinetics, a single
ligand ($B$) molecule has to diffuse among {\em many} competing reactive particles $A$. In addition, it
might be forced to find its target within a specific confining geometry, which in principle can be modeled through a 
collection of reflecting boundaries. Such settings define a genuinely many-body problem,
as the overall flux of ligands is shaped by the mutual screening among all the different reactive boundaries
(the {\em reactive environment}),
known as {\em diffusive interaction}~\cite{Traytak:1992fi,Traytak:1995fk,Eun:2013cr}.
In the following we show how this kind of problems can be formulated and solved in a rather general form.\\
\indent Let us imagine a reaction of the kind~\eqref{e:reactionred} to be catalyzed   
at  $N+1$ spherical boundaries $\partial \Omega_\alpha$ of radius (encounter distance)
$\sigma_\alpha=R_\alpha+a$, $\alpha=0,1,\dots,N$ arranged in space at positions $\boldsymbol{X}_\alpha$. 
With reference to the Smoluchowski problem, 
this means that we are explicitly relaxing the assumption of vanishing density of the 
reactive centers $A$.
In the most general setting, each sphere can be endowed with an intrinsic
reaction rate constant $k^\ast_\alpha$, that specifies the conversion rate 
from encounter complex to product at its surface.  
Then, the stationary density of $B$ molecules is the solution of the following
boundary value problem
\begin{subequations}
\begin{align}
&\bigtriangledown^2 u = 0                                              \label{e:BP1} \\
&\left.\left(
       \sigma_\alpha\frac{\partial u}{\partial r_\alpha} 
       - h_{\alpha} u
  \right)\right|_{\partial \Omega_\alpha}=0 \qquad  \alpha=0,1,\dots,N \label{e:BP2} \\
&\lim_{r\to \infty} u = 1                                              \label{e:BP3}
\end{align}
\end{subequations}
where $h_{\alpha}= k^\ast_{\alpha}/k_{S_\alpha}$ with 
$k_{S_\alpha} = 4 \pi D \sigma_\alpha$.
The boundary conditions (BC)~\eqref{e:BP2} are called {\em radiative} or  Robin boundary conditions.
The limits $h_{\alpha}\to\infty$ and $h_{\alpha}=0$ correspond to perfectly absorbing (sink) 
and reflecting (obstacle) boundaries, respectively, while values $0<h_\alpha<\infty$ 
correspond to finite surface reactivity~\cite{Collins:1949aa}. \\
\indent The  boundary problem~\eqref{e:BP1},~\eqref{e:BP2},~\eqref{e:BP3} provides 
a rigorous mathematical description of a wide assortment of physical situations, 
ranging from one or many sinks screened by neighboring competing reactive boundaries 
to hindered diffusion to a sink located among 
a collection of static reflecting obstacles placed at given positions in space.
In order to solve the problem, it is expedient to consider as many sets 
of spherical coordinate systems as there are boundaries, 
$\boldsymbol{r}_\alpha\equiv(r_\alpha,\theta_\alpha,\phi_\alpha)$. 
The solution can then be written formally as an 
expansion in series of irregular solid harmonics, namely
\begin{equation}
\label{e:ugen}
u = 1 + \sum_{\alpha=0}^N u^-_\alpha(\boldsymbol{r}_\alpha), \qquad
u^-_\alpha =  \sum_{\ell=0}^{\infty}\sum_{m=-\ell}^{\ell}
                    \frac{B_{m\ell}^\alpha} 
                    {r_\alpha^{\ell+1}} 
                    \, Y_{m\ell}(\boldsymbol{r}_\alpha)
\end{equation}
%
%
%\begin{equation}
%u^-_\alpha =  \sum_{\ell=0}^{\infty}\sum_{m=-\ell}^{\ell}
%                    B_{m\ell}^\alpha \, 
%                    r_\alpha^{-(\ell+1)} 
%                    Y_{m\ell}(\boldsymbol{r}_\alpha)
%\label{e:u0a}
%\end{equation}
%
where  $Y_{m\ell}(r_\alpha, \theta_\alpha\phi_\alpha)$ are spherical 
harmonics expressed in the local frame centered at the $\alpha$-th sphere~\footnote{Here 
we use the definition $Y_{m\ell}(\theta,\phi) = P_{\ell}^{m}(\cos\theta) e^{im\phi}$
where $P_{\ell}^{m}(\cos\theta)$ are associated Legendre polynomials~\cite{Morse:1953le}.}.\\
%Here $\alpha=0$ denotes the central sink, while $\alpha=1,2,\dots,N$ label the surrounding screening
%spheres.\\  
%
\indent The coefficients $B_{m\ell}^\alpha$ should be determined by imposing the 
BCs~\eqref{e:BP2}. In order to do so, we use known addition theorems for spherical 
harmonics~\cite{Morse:1953le} to express the solution~\eqref{e:ugen} in all the $N+1$ different reference frames centered 
at each sphere. The result is the following infinite-dimensional system of linear equations
%
%\begin{multline}
\begin{equation}
\label{e:linsysB}
\displaystyle B_{gq}^{\alpha} - \frac{(q-h_\alpha)}{(h_\alpha+q+1)}
               \bigg[
                   \delta_{g0}\delta_{q0}  %\\
                    + \sum_{\ell=0}^{\infty}\sum_{m=-\ell}^{\ell}
                      \sum_{\stackrel{\beta=0}{\beta\neq \alpha}}^{N}B_{m\ell}^{\beta} 
                       W_{m\ell}^{\alpha\beta g q} 
               \bigg] = 0 
\end{equation}
%\end{multline} 
% 
for $\alpha=0,1,\dots,N$, $q=1,2,\dots,\infty$ and $g=-q,...,q$
(see supplementary material for a detailed derivation and the explicit 
expression of the matrix elements $W_{m\ell}^{\alpha\beta g q}$).
% 
%\begin{multline}
%\begin{equation}
%\label{e:Wmat}
%W_{m\ell}^{\alpha\beta g q} = (-1)^{q+g} \frac{(\ell-m+q+g)!}{(\ell-m)!(q+g)!} %\\
%                               \times
%                               \bigg(
%                                       \frac{\sigma_{\alpha}^q \sigma_{\beta}^{\ell+1}}{L_{\beta\alpha}^{\ell+q+1}}
%                               \bigg) Y_{m-g,\ell+q}(\boldsymbol{L}_{\beta\alpha})
%\end{equation}
%\end{multline}
%
%with $\boldsymbol{L}_{\beta\alpha} = \boldsymbol{X}_\beta-\boldsymbol{X}_\alpha$.
%
%is the vector connecting 
%the centers of spheres $\beta$ and $\alpha$
%in the direction $\beta \rightarrow \alpha$ 
%.
To solve the problem one simply needs to truncate the sum on $\ell$ in eq.~\eqref{e:linsysB}, 
by including a finite number of multipoles so as to attain the desired accuracy 
on the overall rate constant. In analogy to eq.~\eqref{e:ratec}, and taking into 
account the definition~\eqref{e:ugen}, the rate constant
corresponding to a given subset of reactive boundaries $\mathcal{S}$ 
can be computed as
\begin{equation}
k = D \sum_{\alpha\in\mathcal{S}}
      \int\limits_{\partial \Omega_\alpha}
      \left. \frac{\partial u}{\partial r} \right|_{r=\sigma_\alpha} dS =
      - \sum_{\alpha\in\mathcal{S}} k_{S_\alpha} B^{\alpha}_{00}
\label{e:ratec1}
\end{equation}
%
%++++++++++++++++++++++++++++++++++++++++++++++++++++++++++++++++++++++++++++++++++++++++++++++++++++++++++++++++++++
%
%                                               R E S U L T S
%
%\section{Results}
%
%++++++++++++++++++++++++++++++++++++++++++++++++++++++++++++++++++++++++++++++++++++++++++++++++++++++++++++++++++++
%%%%%%%%%%%%%%%%%%%%%%%%%%%%%%%%%%%%%%%%%%%%%%%%%%%%%%%%%%%%%%%%%%%%%%%%%%%%%%%%%%%%%%%%%%%%%%%%%%%%%%%%%%%%%%%%%%%%%%%
\begin{figure}[t!]
\centering
\includegraphics[width=\columnwidth]{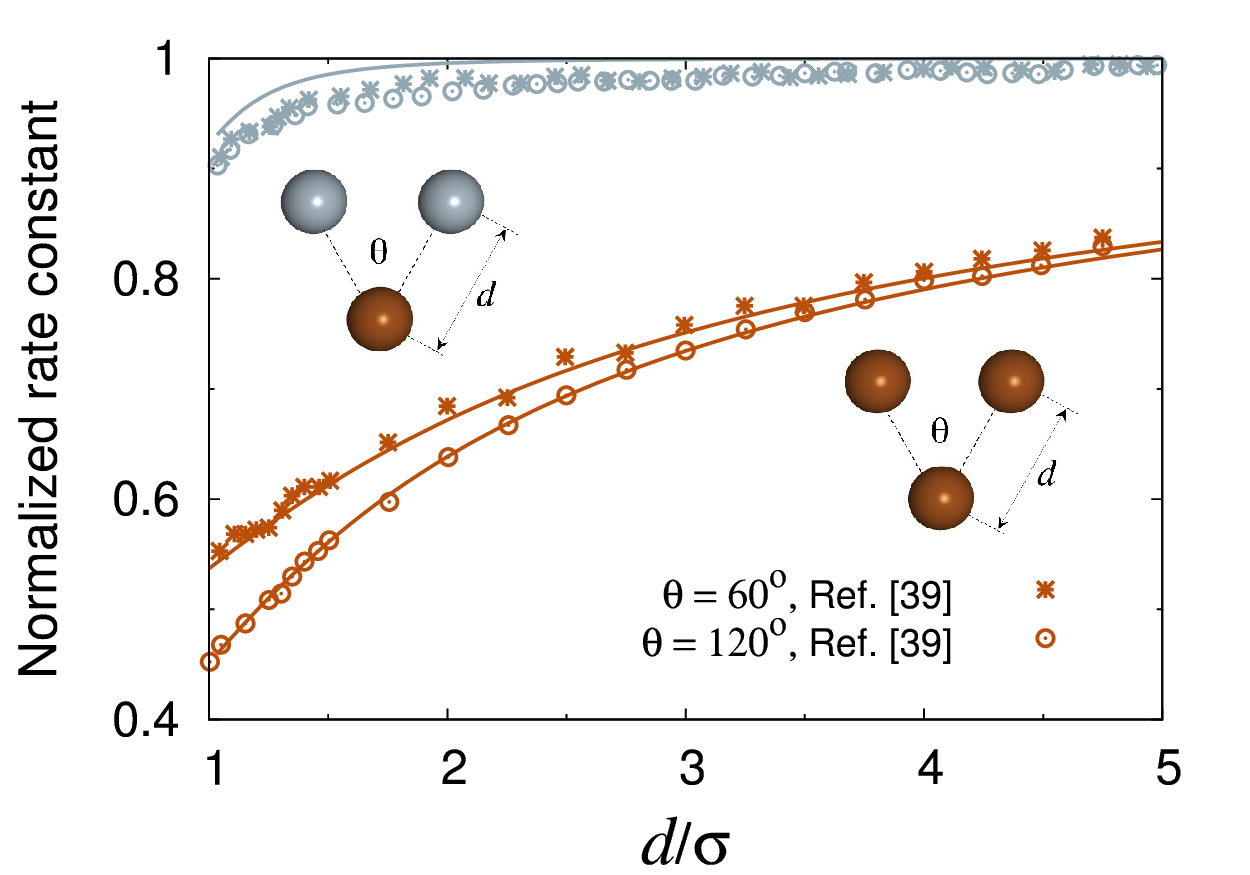}
\caption{\textbf{The approximate finite-element calculations 
compared with the exact results.} Total flux into a sink of diameter $\sigma$ normalized 
to $k_S = 4\pi D\sigma$ (flux into an isolated sink) in the presence of two spherical screening boundaries 
of the same diameter placed at a fixed distance $d$ from the sink and forming an angle $\theta$. 
Light blue and dark orange denote reflecting and absorbing particles, respectively. 
Symbols are numerical results of finite-element 
calculations from Ref.~\cite{Eun:2013cr}. Solid lines are the exact results, 
obtained by solving eqs.~\eqref{e:linsysB} with a relative accuracy of $10^{-4}$.
%
% In this case all particles have the same radius, $R_\alpha=R$ $\forall \ \alpha$
% and $\sigma=a+R$.
}
\label{f:ratemccammon}
\end{figure}
%%%%%%%%%%%%%%%%%%%%%%%%%%%%%%%%%%%%%%%%%%%%%%%%%%%%%%%%%%%%%%%%%%%%%%%%%%%%%%%%%%%%%%%%%%%%%%%%%%%%%%%%%%%%%%%%%%%%%
%
%
%%%%%%%%%%%%%%%%%%%%%%%%%%%%%%%%%%%%%%%%%%%%%%%%%%%%%%%%%%%%%%%%%%%%%%%%%%%%%%%%%%%%%%%%%%%%%%%%%%%%%%%%%%%%%%%%%%%%%%
\begin{figure}[h!]
\centering
\includegraphics[width=\columnwidth]{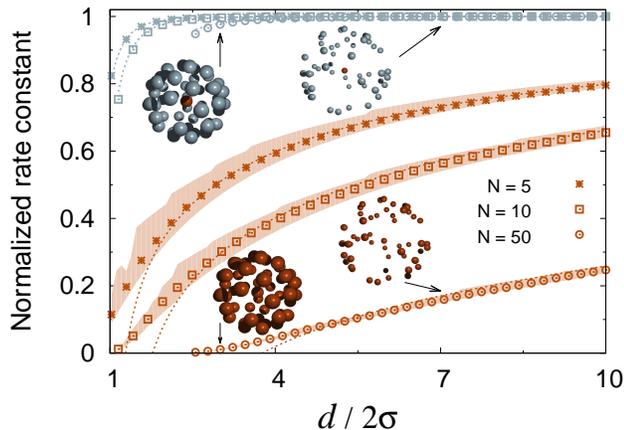}
\caption{\textbf{Competitive screening greatly reduces the rate constant 
compared to inert obstacles, and is strongly modulated by the configuration.} 
Total flux into a sink of radius $\sigma$  surrounded by
$N$ spherical boundaries of the same size arranged randomly at distance $d$
(normalized to $k_S = 4\pi D\sigma$). 
%Light blue and dark orange denote reflecting particles and sinks, respectively. 
Symbols denote the exact results (solution of eqs.~\eqref{e:linsysB}),
averaged over 100 independent configurations for each value of $d$.
The shaded bands highlight the regions comprised between the minimum and maximum rates. 
For reflecting screening boundaries, these regions are as small as the truncation error. 
The light blue and orange lines are plots of eq.~\eqref{e:kNscreen_ref} and 
eq.~\eqref{e:kNscreen_aver}, respectively, with $\lambda=1$. 
The arrows flag values of $d$ corresponding to the two configurations shown ($N=50$) 
with the screening spheres made all absorbing (bottom) and all reflecting (top).}
\label{f:rates}
\end{figure}
%%%%%%%%%%%%%%%%%%%%%%%%%%%%%%%%%%%%%%%%%%%%%%%%%%%%%%%%%%%%%%%%%%%%%%%%%%%%%%%%%%%%%%%%%%%%%%%%%%%%%%%%%%%%%%%%%%%%%
%
\noindent The theoretical framework that culminates in formula~\eqref{e:ratec1} provides an extremely
powerful tool to investigate how specific geometries of obstacles and/or competitive reactive boundaries
modulate the rate constant of a given diffusion-influenced reaction.  \\
%
%
%   UN BUON PUNTO DOVE METTERE UNA SECTION .... per esempio  \section{Diffusion of a ligand to a perfect sink}
%
\indent A clear and instructive illustration of our general approach can be outlined by focussing on the 
simplest model of diffusion-influenced reaction, namely diffusion of a ligand to a perfect sink. 
Even if our method could be employed to examine far more complex {\em reactive geometries},
realized by assembling a large number of spherical boundaries of arbitrary sizes and reactivities, 
for the sake of clarity we shall specialize here to the case of a sink of radius $\sigma$
surrounded by $N$ identical spheres of radius $\sigma_1 = \lambda \sigma$
arranged randomly at a fixed distance $d$. This problem has been tackled recently  
for $\lambda=1$ and $N \leq 4$ through a numerical finite-element (FE) method in Ref.~\cite{Eun:2013cr}. 
This study provided clear-cut hints of the subtle effects brought about by the environment's geometry,
but also highlighted the impossibility of brute-force numerical approaches to assess 
the impact of more crowded and sophisticated reactive environments.\\
%
% The long-range nature of diffusive interactions ~\cite{Traytak:1992fi}. 
% Finite-element computations for 7 sinks in the case of a similar problem arising in heat conduction 
% required about ten times more computational times~\cite{Gordeliy:2009aa} 
%
\indent In Fig.~1 %\ref{f:ratemccammon} 
we compare the FE numerical results with the exact solution for $N=2$. 
It appears clear that the screening effect is harder to capture 
via a FE scheme in the case of reflecting obstacles than in the presence of 
competitive sinks. However, our exact approach allows one to dig much further into this problem and 
investigate analytically the screening effect of configurations comprising a large number of spheres. 
For example, one can expand the system~\eqref{e:linsysB} in powers of $\varepsilon \equiv\sigma/d$ 
to derive simple analytical estimates of the rate constant to the sink (see supplementary material 
for the detailed calculations).  In the case of reflecting obstacles, one gets
%the expansion up to 6-th order reads 
%
\begin{equation}
\label{e:kNscreen_ref}
\frac{k}{k_S} = 1 - \left( 
                       \frac{\lambda^3N}{2} 
                    \right)  \varepsilon^4                   %\bigg(\frac{\sigma}{d}\bigg)^4
                   - \left( 
                       \frac{2\lambda^5N}{3} 
                     \right) \varepsilon^6                   %\bigg(\frac{\sigma}{d}\bigg)^6 
                   + \dots
\end{equation}
which is independent of the screeninig configuration and linear in $N$, 
as suggested in Ref.~\cite{Eun:2013cr}. However, we find that this only holds up to sixth order in 
$\varepsilon$ -- it can be seen from the expansion 
that the configuration enters explicitly successive powers of $\varepsilon$ (see 
supplementary material).  On the other hand, a similar procedure in the case of 
$N$ screening {\em sinks} yields 
%
%\begin{widetext}
\begin{equation}
%\begin{aligned}
\begin{split}
\label{e:dipoloassorbenti}
\frac{k}{k_s} = \, %&
                   1 - \lambda N \varepsilon   %\bigg(\frac{\sigma}{d}\bigg) 
                  + \bigg[
                            \lambda N + \lambda^2\!\!\sum_{\substack{\alpha,\beta=1\\ \beta \neq \alpha}}^N
                                \frac{1}{\Gamma_{\alpha\beta}}
                    \bigg] \varepsilon^2   %\bigg(\frac{\sigma}{d}\bigg)^2 
                    \ \ \ \ \ \ \ \ \ \ \ \ \ \ \ \ \ \ \ \ \ \\
                 %&
                  - \bigg[  
                            \lambda^2 N^2 + \lambda^2\!\!\sum_{\substack{\alpha,\beta=1\\ \beta \neq \alpha}}^N
                                  \frac{1}{\Gamma_{\alpha\beta}}+
                                  \lambda^3\!\!\!\!\sum_{\substack{\alpha,\beta,\delta=1\\ \beta,\delta \neq \alpha}}^N
                                  \frac{1}{\Gamma_{\alpha\beta}\Gamma_{\alpha\delta}}
                    \bigg] \varepsilon^3   %\bigg(\frac{\sigma}{d}\bigg)^3
                    + \dots
%\end{aligned} 
\end{split}                   
\end{equation}
%\end{widetext}
%
where $\Gamma_{\alpha\beta} = 2\sin (\omega_{\alpha\beta}/2)$, 
$\omega_{\alpha\beta}$ being the angle formed by the sinks $\alpha$ and $\beta$
with respect to the origin.  Eq.~\eqref{e:dipoloassorbenti} makes it very clear that the 
configuration of competitive reactive boundaries  does influence the screening effect on the central sink. 
A clear signature of this is also that the corrections in Eq.~\eqref{e:dipoloassorbenti} alternate in sign. 
This observation sheds considerable light onto the many-body character of the rate constant, whose perturbative series 
is alternatively reduced by the diffusive interactions between the screening boundaries and the sink 
(shielding ligand flux from it) and increased by 
the diffusive interactions among the screening particles (shielding flux from each other).
On the contrary, the screening action of inert obstacles is largely dominated by the excluded-volume effect, 
and thus can only yield negative corrections at all orders.\\
\indent Due to its perturbative nature, eq.~\eqref{e:dipoloassorbenti} can be 
used to quantify the shielding action of specific 3$D$ arrangements of sinks
only for  $N \varepsilon \propto N/d \ll 1$. However, it still provides
a powerful analytical tool to {\em compare} different geometries, as the 
perturbative rate is always proportional to the true rate (see supplementary material). 
For example, eq.~\eqref{e:dipoloassorbenti} could be used to 
{\em design} the special configurations that minimize or maximize the screening effect on 
the central sink for given values of $N$ and $d$.\\
\indent The  {\em average} shielding action exerted by $N$ equidistant sinks 
can be easily obtained analytically in the monopole approximation, {\em i.e.}
by keeping only the $\ell=0$ and $q=0$ terms in eqs.~\eqref{e:linsysB}.
The ensuing reduced system can be averaged over different configurations 
in the hypothesis of vanishing many-body spatial correlations, {\em i.e.}
by integrating over the probability density
$\mathcal{P}_N = \prod_{\alpha\neq\beta}(\sin\omega_{\alpha\beta})/4\pi$,
with $2\arcsin(\sigma_1/d)\leq\omega_{\alpha\beta}\leq\pi$ (excluded-volume 
constraint between screening sinks). The result is (see supplementary material for the details)
\begin{equation}
\label{e:kNscreen_aver}
\left\langle 
      \frac{k}{k_S} 
\right\rangle = 
      \frac{1 - \lambda\varepsilon [N - (N-1)(1-\lambda\varepsilon) ]}
           {1 - \lambda\varepsilon [N\varepsilon - (N-1)(1-\lambda\varepsilon) ]}
\end{equation}
Fig.~\ref{f:rates} 
shows that for $\lambda=1$ eq.~\eqref{e:kNscreen_aver} provides 
an extremely good estimate of the configurational averages of the exact results 
at separations greater than a 
few diameters, % of the central sink, 
highlighting the dramatic screening action of 
competitive reactive boundaries with respect to inert obstacles. 
Furthermore, a simple analysis of the rate {\em fluctuations} over the configuration ensembles at fixed $d$
allows one to gauge how sensitive competitive screening 
is to the specific 3$D$ arrangement of the sinks. 
Remarkably, this analysis reveals stretches between the minimum and maximum 
rates for a given value of $d$ as high as 50 \% of the average
(see shaded bands in Fig.~\ref{f:rates}). 
More precisely, we remark that the variability associated with different 
geometries is greater ($i$) at short distances and ($ii$) for few screening particles.\\
%(see again  Fig.~2 %\ref{f:rates}
%).  
%
%We note that for $N=1$ eq.~\eqref{e:kNscreen_aver} reduces to the known
%formula for a system of two identical sinks~\cite{Traytak:1992fi}, $k/k_S=1/(1+\varepsilon)$.
%
%%%%%%%%%%%%%%%%%%%%%%%%%%%%%%%%%%%%%%%%%%%%%%%%%%%%%%%%%%%%%%%%%%%%%%%%%%%%%%%%%%%%%%%%%%%%%%%%%%%%%%%%%%%%%%%%%%%%%
\begin{figure}
\includegraphics[width=\columnwidth]{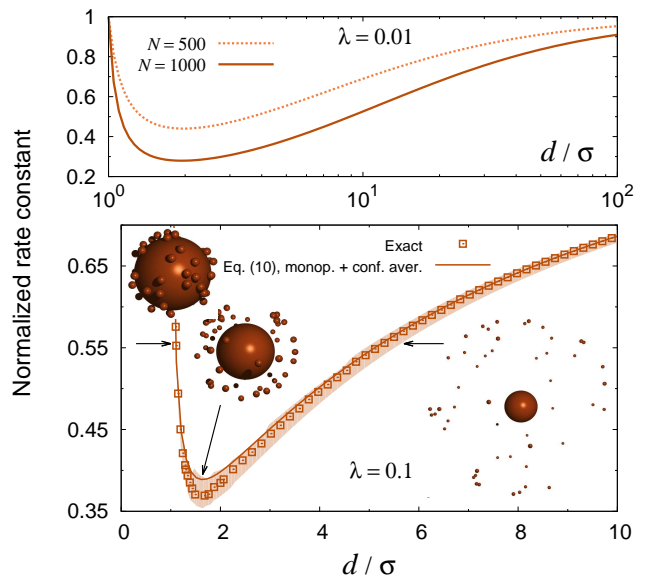}
\caption{\textbf{Making the screening boundaries smaller makes
the flux into the sink non-monotonic}. Total flux into a sink of radius 
$\sigma$ surrounded by $N=50$ smaller sinks of radius $\sigma_1=\sigma/10$ arranged randomly 
at a distance $d$ (normalized to $k_S = 4\pi D\sigma$). 
The left-most and right-most cartoons depict two configurations that screen exactly the same amount of flux, 
despite being at considerably different distances  ($d/\sigma = 1.1$ and $d/\sigma = 8$). 
The configuration shown in the middle corresponds to the 
predicted minimum at $d/\sigma = 1+\sqrt{1 - \lambda[1+(N-1)\lambda]}\approx 1.64$.
The solid line is a plot of formula~\eqref{e:kNscreen_aver}. Each symbol is the average over 250
independent configurations, while the filled band comprises the region between the minimum and maximum rates. 
The top panel illustrates the case of screening by a large number of tiny particles, 
highlighting the sizeable non-monotonic effect. The curves are plots of eq.~\eqref{e:kNscreen_aver}.
\label{f:rateLS}}
\end{figure}
%%%%%%%%%%%%%%%%%%%%%%%%%%%%%%%%%%%%%%%%%%%%%%%%%%%%%%%%%%%%%%%%%%%%%%%%%%%%%%%%%%%%%%%%%%%%%%%%%%%%%%%%%%%%%%%%%%%%%
%
\indent Eqs.~\eqref{e:kNscreen_ref},~\eqref{e:dipoloassorbenti} and~\eqref{e:kNscreen_aver}
and the ensuing arguments are rather exemplary illustrations of the powerful analytical insight afforded 
by our general approach.  The case of small screening sinks, $\lambda <1$, provides a further demonstration 
of non-trivial effects that are captured by our analytics. It turns out that 
the function~\eqref{e:kNscreen_aver} displays a minimum for certain 
choices of the parameters $N,\lambda$. More precisely, 
a minimum exists at fixed $N$ for $\lambda \leq \lambda^\ast(N) \equiv (\sqrt{4N+1}-1)/(2N) < 1$, 
or, alternatively, at  fixed $\lambda$ for  $N\leq N^\ast(\lambda)\equiv (1-\lambda)/\lambda^2$.
Fig.~\ref{f:rateLS} provides a clear illustration of this subtle effect.\\
\indent The flux into a large sink features 
a minimum for screening configurations of tiny absorbing 
particles close to its surface. This is the result of the 
competition between two effects. 
When the small particles lie very close to the surface of the large sink $\mathcal{S}_\sigma$,
the latter behaves as an {\em effective isolated} sink of size $\sigma_{\rm eff}<\sigma$, 
absorbing a flux $\Phi_{\rm eff} = 4\pi D\sigma_{\rm eff} \lesssim k_S$.
Upon increasing the distance $d$, the total flux to the 
screening sinks will increase (their {\em active} surfaces get larger and 
they also get farther apart from each other). Now it is clear that the effect of this 
on the flux to $\mathcal{S}_\sigma$ will depend on the size of the screening particles.
If $\sigma_1$ is small enough, $\sigma_{\rm eff}$ is not much smaller than $\sigma$, 
so that $\Phi_{d=\sigma+\sigma_1}\equiv\Phi_{\rm eff}$ is not much smaller than 
$\Phi_\infty=k_S$. 
Under these conditions, the flux into the large sink starts {\em decreasing},
as the screening ensemble effectively {\em steal} more and more flux from it. 
However, upon increasing $d$ past a critical distance,
the small particles can no longer catch enough ligand flux,  
so that the flux to $\mathcal{S}_\sigma$ starts increasing, as it should, towards $k_S$.\\
%
%On the one hand, the flux decreases as the screening 
%spheres get closer, since the diffusive interactions between them and the sink increase (each 
%of them shields more and more flux). On the other hand, as they approach, the 
%diffusive interactions {\em among them} also increase, which has the opposite effect 
%of globally shielding less flux from the large sink (they shield more and more flux 
%from each other). Past a critical distance, where the latter effect takes over if the 
%screening sinks are small enough, 
%the rate starts increasing. 
%
%
%\indent The complex nature of this many-body kinetics is also 
%evident from the fluctuations associated with different ensembles of configurations 
%at a given value of $d$. The width of the shaded band in Fig.~3 %\ref{f:rateLS} 
%displays a clear maximum (increased configurational variability) where the rate constant is at the 
%stationary point, spotlighting the apex of the interplay between the two 
%aforementioned opposite influences on the rate constant.
%These are additional striking demonstrations of the kind of analytical insight afforded by
%our method to investigate how the geometrical features of the environment fine-tunes the 
%rate constant of diffusion-influenced  chemical reactions.\\
%
\indent Summarizing, in this paper we have introduced a general theoretical framework to quantify 
how the geometry and distributed reactivity patterns of the environment modulate the rate constant 
of diffusion-influenced chemical reactions. Our method can be used to examine arbitrary 
{\em reactive landscapes}, made by assembling spherical boundaries of selected size at given locations 
in space and endowed with arbitrary surface reactivity. We note that our method is utterly general, 
as it can be easily extended to accommodate for reactive environments realized with  
more complex, non-spherical boundaries. The only requirement is that one of the 13 coordinate systems 
in which Laplace's equation is separable be used~\cite{Morse:1953le}, 
and that addition theorems exist for the corresponding elementary solutions~\cite{Hobson:1931aa}.
Moreover,  our method can be extended to Laplace space~\cite{Gordeliy:2009aa}, so as to work 
out exactly the effect of the environment on {\em time-dependent} problems. This technique could be 
employed to shed further light on the intriguing sensitivity of time-dependent effects
on initial conditions, which seems to constitute a rather generic feature of complex 
media~\cite{Benichou:2010aa}.\\
\indent Finally, we stress that our method can be easily employed to derive approximate closed 
analytic formulas, that can be used to investigate naturally occurring reactive geometries and 
to assist in the design of effective artificial nano-reactors for different technological 
applications.

%
% ACKNOWLEDGEMENTS
%

%\acknowledgements
\smallskip
\noindent The authors wish to thank Sergey Traytak, P. De Los Rios and G. Foffi 
for insightful comments. F. P. and D. F. acknowledge joint funding from the French CNRS (PICS). 

%project ``Theoretical models of 
%diffusion in crowded and confining media: applications to intracellular transport (2014-2016)''
 
%====================================================================================================================
%
%  BIBLIOGRAPHY
%

%% Put the bibliography here, most people will use BiBTeX in
%% which case the environment below should be replaced with
%% the \bibliography{} command.

%% \bibliography{bio-11,DFT-FMT,tesi,bib_igg,crowding,diff_obstacles,Refs,nanoreactors}


\end{document}